\documentclass[preprint,preprintnumbers, prd, floatfix,  superscriptaddress,nofootinbib] {revtex4-1}
\usepackage{graphicx}
\usepackage{epsfig}
\usepackage{bm}
\usepackage{amssymb}
\usepackage{float}
\usepackage{amsmath}
\usepackage{dcolumn}
\usepackage{cancel}
\usepackage[colorlinks]{hyperref}
\usepackage[usenames,dvipsnames]{color}
\usepackage{color}
\usepackage{epstopdf}
\usepackage{nccbbb}

\newcommand{\CD}{{\cal D}}
\newcommand{\CR}{{\cal R}}
\newcommand{\CQ}{{\cal Q}}

\newcommand{\average}[1]{\left\langle #1 \right\rangle_\CD}

\newcommand{\naverage}[1]{\left\langle #1 \right\rangle_{\CD_{\rm \bf 0}}}

\newcommand{\now}[1]{{#1_{\rm \bf 0}}}
\newcommand{\be}{\begin{equation}}
\newcommand{\ee}{\end{equation}}
\newcommand{\bea}{\begin{eqnarray}}
\newcommand{\eea}{\end{eqnarray}}
\newcommand{\bean}{\begin{eqnarray*}}
\newcommand{\eean}{\end{eqnarray*}}
\hypersetup{
     breaklinks=true,
    pdfstartview={FitH},    colorlinks=true,       % false: boxed links; true: colored links
    linkcolor=blue,          % color of internal links
    citecolor=red,        % color of links to bibliography
    filecolor=magenta,      % color of file links
    urlcolor=blue,           % color of external links
    anchorcolor=green,      % Color for anchor text
    linktocpage=true
}

\begin{document}
\title{CPL effective dark energy from the backreaction effect}

\author{Yan-Hong Yao}
\email{yhy@mail.nankai.edu.cn}
\author{Xin-He Meng}
\email{xhm@nankai.edu.cn}

\affiliation{Department of Physics, Nankai University, Tianjin 300071, China}

\begin{abstract}
In this paper, we interpret the dark energy as an effect caused by small scale inhomogeneities of the universe with the use of the spatial averaged approach of Buchert\cite{buchert2000average,buchert2001average}. The model considered here adopts the Chevallier-Polarski-Linder(CPL) parameterizations of the equation of state of the effective perfect fluid  from the backreaction effect. Thanks to the effective geometry introduced by Larena et. al.\cite{larena2009testing} in their previous work, we confront such backreaction model with latest type Ia supernova and Hubble parameter observations, coming out with results that reveal the difference between the Friedmann-Lema\^{\i}tre-Robertson-Walker model and backreaction model.

\end{abstract}

\maketitle

\section{Introduction}
\label{intro}
According to recent observations of type Ia supernovae, the universe is in a state of accelerated expansion \cite{riess1998,perlmutter1999measurements}, and the simplest scenario to account for these observations is the so called Lambda cold dark matter $(\Lambda CDM) $ model, which is also named standard cosmology which inclues the simplest dark energy model. However, because of the huge discrepancy between the theoretical expected value of the cosmological constant and the observed one, other alternative scenarios have been proposed, including scalar field models such as quintessence\cite{Caldwell1998Cosmological}, phantom\cite{Caldwell1999A}, dilatonic\cite{Piazza2004Dilatonic}, tachyon\cite{Padmanabhan2002Accelerated}and quintom\cite{Bo2006Oscillating} etc. and modified gravity models such as braneworlds \cite{maartens2010brane}, scalar-tensor gravity \cite{esposito2001scalar}, higher-order gravitational theories \cite{capozziello2005reconciling,das2006curvature}. Whereas the so called fitting problem that how well is our universe described by a standard Friedmann-Lema\^{\i}tre-Robertson-Walker(FLRW) model is not solved yet, recently a third alternative has been considered to explain the dark energy phenomenon as backreaction effect(a Large scale effect caused by small scale inhomogeneities of the universe \cite{rasanen2004dark,kolb2006cosmic}.).

In order to consider backreaction effect, it is necessary to answer a longstanding question that how to average a general inhomogeneous spacetime. To date the macroscopic gravity (MG) approach \cite{zalaletdinov1992averaging,zalaletdinov1997averaging,zalaletdinov1993towards,mars1997space} is the only approach that gives a prescription for the correlation functions which emerge in an averaging of the Einstein's field equations, however, so far it required a number of assumptions about the correlation functions, which make the theory less convictive. Therefore, in this paper we adopt another averaged approach put forward by Buchert \cite{buchert2000average,buchert2001average}, in despite of its foliation dependent nature, such approach is quite simple and hence becomes the most well studied averaged approach. Since the final averaged equations in such approach do not form a closed set, one needs to make some assumptions about the backreaction term appeared in the averaged equations in order to track the averaged evolution of the universe with certain initial condition. In \cite{buchert2006correspondence}, Buchert
proposes a backreaction model by taking the assumption that the backreaction term ${\cal Q}_\CD$ and the averaged spatial Ricci scalar $\average{\CR}$ obey the scaling laws of the volume scale factor $a_\CD$. Although simple, such model is able to describe a accelerated universe without introducing strange negative pressure matter. Of course, there is no conclusive evidences that scaling laws assumption is absolutely correct, thus other supposition about the behaviors of the backreaction term can be make. In this paper, we propose a new backreaction model by assuming that the equation of state of the effective perfect fluid follows the CPL form and using a relation appeared in \cite{QING2013CONSTRAINTS}  to reduce a parameter for simplicity reason. after then, we apply the template metric presented in \cite{buchert2006correspondence} to confront the new model with observations and perform a likelihood analysis. We use the natural unit c=1 through the paper.

The paper is organized as follows. In Section \ref{sec:1}, the spatial averaged approach of Buchert is demonstrated with presentation of the averaged equations for the volume scale factor $a_\CD$. In Section \ref{sec:2}, we introduce the template metric, which is a necessary tool to test the theoretical preditions with observations, and computation of observables. In Section \ref{sec:3}, we apply a likelihood analysis of the backreaction models by confronting it with latest type Ia supernova and Hubble parameter observations. After analysis of the results in Section \ref{sec:3}, we summarize ourresults in the last section.

\section{The backreaction models}
\label{sec:1}
In \cite{buchert2000average}, Buchert considers a universe filled with irrotational dust with energy density $\varrho$. By foliating space-time with the use of Arnowitt-Deser-Misner(ADM) procedure and defining an averaging operator that acts on any spatial scalar $\Psi$ function as
\begin{equation}
\average{\Psi}:=\frac{1}{V_{\CD}}\int_{\CD}\Psi Jd^{3}X,
\end{equation}
where $V_{\CD}:=\int_{\CD}Jd^{3}X$ is the domain's volume, Buchert obtains two averaged equations here we need, the averaged Raychaudhuri equation
\begin{equation}
3\frac{{\ddot a}_\CD}{a_\CD} + 4\pi G \average{\varrho} ={\CQ}_\CD,
\end{equation}
and the averaged Hamiltonian constraint
\begin{equation}
3\left( \frac{{\dot a}_\CD}{a_\CD}\right)^2 - 8\pi G \average{\varrho}= - \frac{\average{\CR}+{\CQ}_\CD }{2}.
\end{equation}
In these two equations, $a_\CD(t) = \left(\frac{V_\CD (t)}{V_{\now\CD}}\right)^{1/3}$ is the volume scale factor, where $V_{\now\CD} =\vert{\now\CD}\vert$ denotes the present value of the volume, and ${\cal Q}_\CD$, $\average{\CR}$  represent the backreaction term and the averaged spatial Ricci scalar respectively, which are related by the following integrability condition
\begin{equation}
\frac{1}{a_\CD^6}\partial_t \left(\,{\CQ}_\CD \,a_\CD^6 \,\right)
\;+\; \frac{1}{a_\CD^{2}} \;\partial_t \left(\,\average{\CR}a_\CD^2 \,
\right)\,=0\;.
\end{equation}
Now we can define the effective prefect fluid by
\begin{eqnarray}
  \varrho_{eff}^{\CD}:&=& -\frac{1}{16\pi G}({\CQ}_\CD+\average{\CR}) \\
  p_{eff}^{\CD}:&=&  -\frac{1}{16\pi G}({\CQ}_\CD-\frac{\average{\CR}}{3})
\end{eqnarray}
 the averaged Raychaudhuri equation and the averaged Hamiltonian constraint then can formally be recast into standard Friedmann equations for a total perfect fluid energy momentum tensor
 \begin{equation}\label{7}
   3\frac{{\ddot a}_\CD}{a_\CD} + 4\pi G (\average{\varrho}+\varrho_{eff}^{\CD}+3p_{eff}^{\CD} )= 0
 \end{equation}
 \begin{equation}\label{8}
    3\left( \frac{{\dot a}_\CD}{a_\CD}\right)^2 = 8\pi G (\average{\varrho}+\varrho_{eff}^{\CD})
 \end{equation}
given the effective energy density and pressure, the effective equation of state reads
\begin{equation}\label{}
w_{eff}^{\CD}:=\frac{p_{eff}^{\CD}}{\varrho_{eff}^{\CD}}=\frac{{\CQ}_\CD-\frac{\average{\CR}}{3}}{{\CQ}_\CD+\average{\CR}}
\end{equation}
One can then obtain a specific backreaction model with an extra ansatz about the form of $w_{eff}^{\CD}$, for example, in this paper, we assume that effective equation of state follows the Chevallier-Polarski-Linder(CPL)form, i.e.
\begin{equation}\label{}
  w_{eff}^{\CD}=w_{0}^{\CD}+w_{a}^{\CD}(1-a_\CD)
\end{equation}
so the Eq.~(\ref{8}) can be rewrite as
\begin{equation}\label{}
  H_{\CD}^{2}=H_{\now\CD}^{2}[\Omega_{m}^{\now\CD}a_{\CD}^{-3}+(1-\Omega_{m}^{\now\CD})a_{\CD}^{-3(1+w_{0}^{\CD}+w_{a}^{\CD})}e^{-3w_{a}^{\CD}(1-a_{\CD})}]
\end{equation}
here $H_\CD : = {\dot a}_\CD / a_\CD$ denotes the volume Hubble parameter and $\Omega_{m}^{\now\CD}:= \frac{8\pi G}{3 H_{\now\CD}^2} \naverage{\varrho}$.
Combine the above equations, we have the following formulas for the backreaction term and the averaged spatial Ricci scalar
\begin{equation}\label{}
{\cal Q}_\CD=-\frac{9}{2}(1-\Omega_{m}^{\now\CD})H_{\now\CD}^{2}e^{-3w_{a}^{\CD}(1-a_{\CD})}[(\frac{1}{3}+w_{0}^{\CD}+w_{a}^{\CD})a_{\CD}^{-3(1+w_{0}^{\CD}+w_{a}^{\CD})}-w_{a}^{\CD}a_{\CD}^{-2-3(w_{0}^{\CD}+w_{a}^{\CD})}]
\end{equation}
\begin{equation}\label{}
  \average{\CR}=-\frac{9}{2}(1-\Omega_{m}^{\now\CD})H_{\now\CD}^{2}e^{-3w_{a}^{\CD}(1-a_{\CD})}[(1-w_{0}^{\CD}-w_{a}^{\CD})a_{\CD}^{-3(1+w_{0}^{\CD}+w_{a}^{\CD})}+w_{a}^{\CD}a_{\CD}^{-2-3(w_{0}^{\CD}+w_{a}^{\CD})}]
\end{equation}
In \cite{QING2013CONSTRAINTS}, Qing Gao and Yungui Gongand show that quinteesence dark energy model can be approximated as the CPL parametrization at low redshift with a degeneracy relationship, here we use this relationship to reduce a parameter in our backreaction model,i.e. we assume
\begin{equation}\label{}
  w_{a}^{\CD}=6(1+w_{0}^{\CD})\frac{(1-\Omega_{m}^{\now\CD})^{-\frac{1}{2}}-(1-\Omega_{m}^{\now\CD})^{\frac{1}{2}}-((1-\Omega_{m}^{\now\CD})^{-1}-1)tanh^{-1}((1-\Omega_{m}^{\now\CD})^{\frac{1}{2}})}{(1-\Omega_{m}^{\now\CD})^{-\frac{1}{2}}-((1-\Omega_{m}^{\now\CD})^{-1}-1)tanh^{-1}((1-\Omega_{m}^{\now\CD})^{\frac{1}{2}})}
\end{equation}
so there are only two parameters($\Omega_{m}^{\now\CD}$ and $w_{0}^{\CD}$ )remain in the model we propose.

\section{Effective geometry}
\label{sec:2}
\subsection{The template metric}
In \cite{larena2009testing}, a template metric was proposed by Larena et. al. as follows,
\begin{equation}
\label{eq:tempmetric1}
{}^4 {\bf g}^\CD = -dt^2 + L_{H_{D_0}}^2 \,a_D^2 \gamma^\CD_{ij}\,dX^i \otimes dX^j \;\;,
\end{equation}
where $L_{\now H}=1/H_{\now\CD}$ is the present size of the horizon introduced so that the coordinate distance is dimensionless,
and the domain-dependent effective three-metric reads:
\begin{equation}
\gamma^\CD_{ij}\,dX^i \otimes dX^j =\frac{dr^2}{1-\kappa_{\CD}(t)r^2}+r^2d\Omega^{2}
\end{equation}
with $d\Omega^{2}=d\theta^{2}+\sin^{2}(\theta)d\phi^2$, this effective three-template metric is identical to the spatial part of a FLRW metric at any given time, but its scalar curvature  $\kappa_{\CD}$  can vary from time to time. As was pointed out by Larena et. al., $\kappa_{\CD}$ cannot be arbitrary, more precisely, they argue that it should be related to the true averaged scalar curvature $\average{\CR}$ in the way that
\begin{equation}
\average{\CR}=\frac{\kappa_{\CD}(t)|\naverage{\CR}|a_{\now\CD}^{2}}{a_{\CD}^{2}(t)}
\end{equation}

\subsection{Computation of observables}
The computation of effective distances along the light cone defined by the template metric is very different from that of distances in FLRW models. Firstly, let us introduce an effective redshift $z_{\CD}$ defined by
\begin{equation}
1+z_{\CD}:=\frac{(g_{ab}k^{a}u^{b})_{S}}{(g_{ab}k^{a}u^{b})_{O}}\mbox{ ,}
\end{equation}
where the letters O and S denote the evaluation of the quantities at the observer and at the source respectively, $g_{ab}$ in this expression represents the template metric, while $u^{a}$ is the four-velocity of the dust which satisfies $u^{a}u_{a}=-1$, $k^{a}$ the wave vector of a light ray travelling from the source S towards the observer O with the restrictions $k^{a}k_{a}=0$. Then, by normalizing this wave vector such that $(k^{a}u_{a})_{O}=-1$ and introducing the scaled vector $\hat{k}^{a}=a_{\CD}^{2}k^{a}$, we have the following equation:
\begin{equation}
\label{eq:defred2}
1+z_{\CD}=(a_{\CD}^{-1}\hat{k}^{0})_{S}\mbox{ ,}
\end{equation}
with $\hat{k}^{0}$ obeying the null geodesics equation $k^{a}\nabla_{a}k^{b}=0$ which leads to
\begin{equation}
\label{eq:evolk}
\frac{1}{\hat{k}^{0}}\frac{d\hat{k}^{0}}{da_{\CD}}=-\frac{r^{2}(a_{\CD})}{2(1-\kappa_{\CD}(a_{\CD})r^{2}(a_{\CD}))}\frac{d\kappa_{\CD}(a_{\CD})}{da_{\CD}}\mbox{ .}
\end{equation}

As usual, the coordinate distance can be derived from the equation of radial null geodesics:
\begin{equation}
\label{eq:coorddist}
\frac{dr}{da_{\CD}}=-\frac{H_{\now\CD}}{a_{\CD}^{2}H_{\CD}(a_{\CD})}\sqrt{1-\kappa_{\CD}(a_{\CD})r^{2}}
\end{equation}

Solving these two equations with the initial condition $\hat{k}^{0}(1)=1, r(1)=0 $ and then plugging $\hat{k}^{0}(a_\CD)$ into Eq.~(\ref{eq:defred2}), one finds the relation between the redshift and the scale factor. With these results, we can determine the volume Hubble parameter $H_{\CD}(z_{\CD})$ and the luminosity distance $d_{L}(z_{\CD})$ of the sources defined by the following formula
\begin{eqnarray}
\label{eq:distances}
d_{L}(z_{\CD})&=&\frac{1}{H_{\now\CD}}(1+z_{\CD})^{2}a_{\CD}(z_{\CD})r(z_{\CD}).
\end{eqnarray}
Having computed these two observables , it is then possible to compare the backreaction model predictions with type Ia supernova and Hubble parameter observations.
\section{Constraints from supernovae data and OHD}
\label{sec:3}
In this section, we perform a likelihood analysis on the parameters of the backreaction model mentioned above with the combination of datasets from type Ia supernova and Hubble parameter observations.
The constraints placed by the recently released Union2.1\cite{Suzuki2012The} compilation on $(\Omega_{m}^{\now\CD},w_{0}^{\CD})$ can be obtained by minimizing
\begin{equation}\label{}
  \chi_{SN Ia}^{2}(\Omega_{m}^{\now\CD},w_{0}^{\CD})= R-\frac{S^{2}}{T}
\end{equation}
here $R$, $S$ and $T$ are defined as
\begin{eqnarray}
  R &=& \sum_{i=0}^{580}\frac{\left(5\log_{10}\left[H_{\now\CD}d_{L}({z_{\CD}}_{i})\right]-\mu_{obs}({z_{\CD}}_{i})\right)^{2}}{\sigma_{\mu}^{2}({z_{\CD}}_{i})},\\
  S &=& \sum_{i=0}^{580}\frac{\left(5\log_{10}\left[H_{\now\CD}d_{L}({z_{\CD}}_{i})\right]-\mu_{obs}({z_{\CD}}_{i})\right)}{\sigma_{\mu}^{2}({z_{\CD}}_{i})}, \\
  T &=& \sum_{i=0}^{580}\frac{1}{\sigma_{\mu}^{2}({z_{\CD}}_{i})}.
\end{eqnarray}

where $\mu_{obs}$ represents the observed distance modulus and $\sigma_{\mu}$ denotes its statistical uncertainty.

For the observed Hubble parameter dataset in Table \ref{tab:1}, the best-fit values of the parameters $(H_{\now\CD},\Omega_{m}^{\now\CD},w_{0}^{\CD})$ can be determined by a likelihood analysis based on the calculation of
\begin{equation}
 \chi^{2}_H(H_{\now\CD},\Omega_{m}^{\now\CD},w_{0}^{\CD})=\sum_{i=0}^{30} \frac{(H_{\CD}({z_{\CD}}_{i};H_{\now\CD},\Omega_{m}^{\now\CD},w_{0}^{\CD})-H_{obs}({z_{\CD}}_{i}))^2}{\sigma_{H}^{2}({z_{\CD}}_{i})}.
\end{equation}
As Ma et. al.\cite{ma2011power} stated, the marginalized probability density function determined by integrating $ \rm e^{-\frac{\chi^{2}_H(H_{\now\CD},\Omega_{m}^{\now\CD},w_{0}^{\CD})}{2}} $ over $H_{\now\CD}$ from $x$ to $y$  with a Gaussian prior reads
\begin{equation}
  {\rm e^{-\frac{\chi^{2}_H(\Omega_{m}^{\now\CD},w_{0}^{\CD})}{2}}} = \frac{1}{\sqrt{A}}[erf(\frac{B}{\sqrt{A}})+1]e^{\frac{B^{2}}{A}}
\end{equation}
where
\begin{equation}
  A=\frac{1}{2\sigma_{H}^{2}}\sum_{i=0}^{30}\frac{H_{\CD}^2({z_{\CD}}_{i};H_{\now\CD},\Omega_{m}^{\now\CD},w_{0}^{\CD})}{2H_{\now\CD}^2\sigma_{H}^{2}({z_{\CD}}_{i})}, B=\frac{\mu_{H}}{2\sigma_{H}^{2}}\sum_{i=0}^{30}\frac{H_{\CD}({z_{\CD}}_{i};H_{\now\CD},\Omega_{m}^{\now\CD},w_{0}^{\CD})H_{obs}({z_{\CD}}_{i})}{2H_{\now\CD}\sigma_{H}^{2}({z_{\CD}}_{i})},
\end{equation}

Finally, the total $\chi^{2}(\Omega_{m}^{\now\CD},w_{0}^{\CD})$ for the combined observational dataset are given by
$\chi^{2}(\Omega_{m}^{\now\CD},w_{0}^{\CD})=\chi_{SN Ia}^{2}(\Omega_{m}^{\now\CD},w_{0}^{\CD})+\chi_{H}^{2}(\Omega_{m}^{\now\CD},w_{0}^{\CD})$.

\begin{figure}
%\begin{tabular}{cc}
\begin{flushleft}
\begin{minipage}{0.45\linewidth}
  \centerline{\includegraphics[width=1\textwidth]{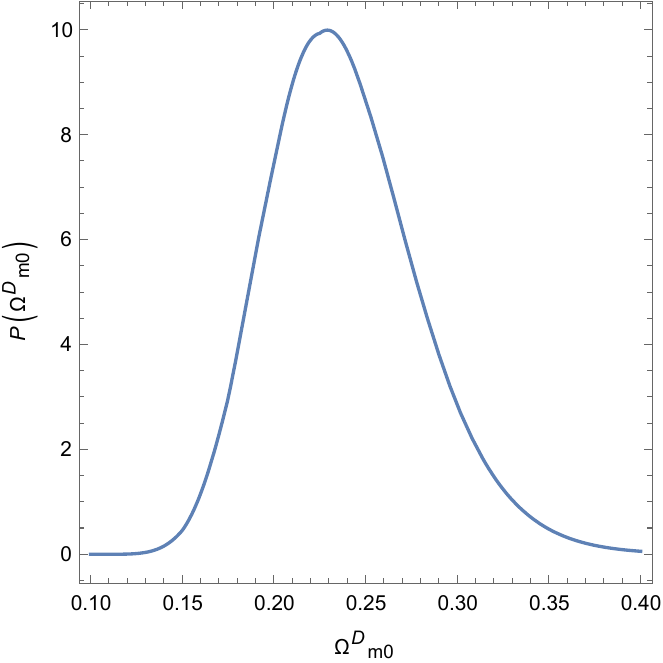}}
\end{minipage}
\hfill
\end{flushleft}

\begin{flushleft}
\begin{minipage}{0.45\linewidth}
  \centerline{\includegraphics[width=1\textwidth]{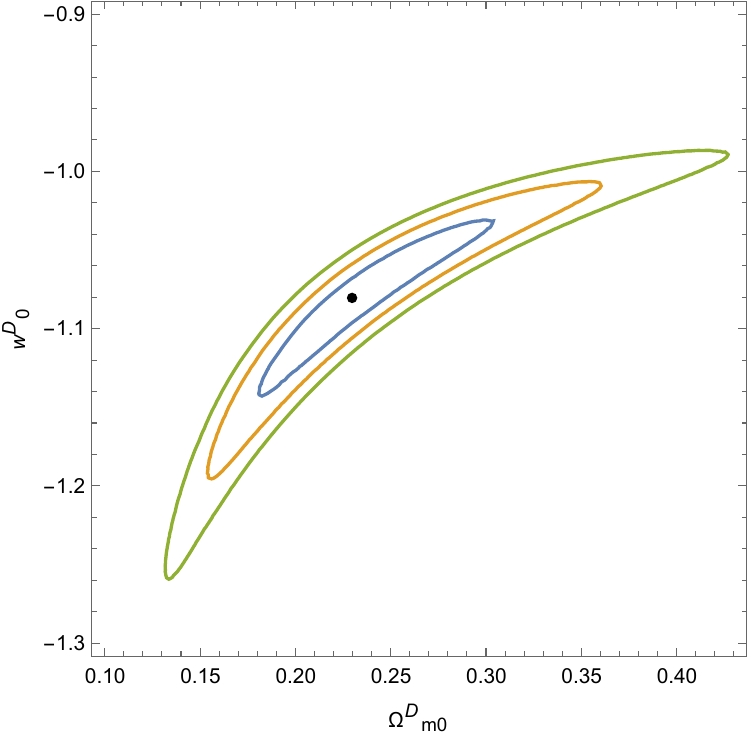}}
\end{minipage}
\hfill
\begin{minipage}{0.44\linewidth}
 \centerline{\includegraphics[width=1\textwidth]{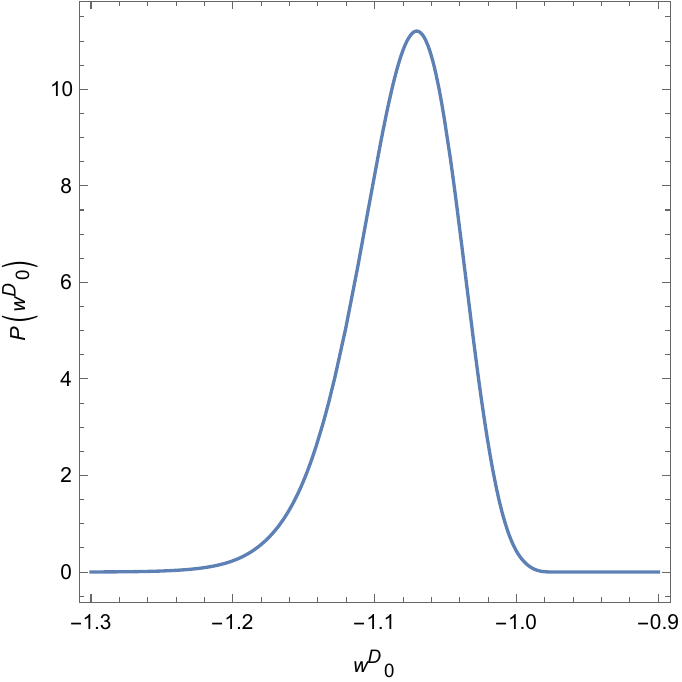}}
\end{minipage}
\end{flushleft}

\caption{The best-fit point and $1\sigma$, $2\sigma$, $3\sigma$ confidence regions of the parameters $\Omega_{m}^{\now\CD},w_{0}^{\CD}$ for this backreaction model, along with their own probability density function.
 The prior for $\Omega_{m}^{\now\CD}\varpropto $ H(0.4-$\Omega_{m}^{\now\CD}$)H($\Omega_{m}^{\now\CD}$-0.1),
The prior for $w_{0}^{\CD}\varpropto$ H(-0.9-$w_{0}^{\CD}$)H($w_{0}^{\CD}$-(-1.3)),
H(x)is the step function.
}
\label{fig:1}
\end{figure}

\begin{figure}
%\begin{tabular}{cc}
\begin{flushleft}
\begin{minipage}{0.45\linewidth}
  \centerline{\includegraphics[width=1\textwidth]{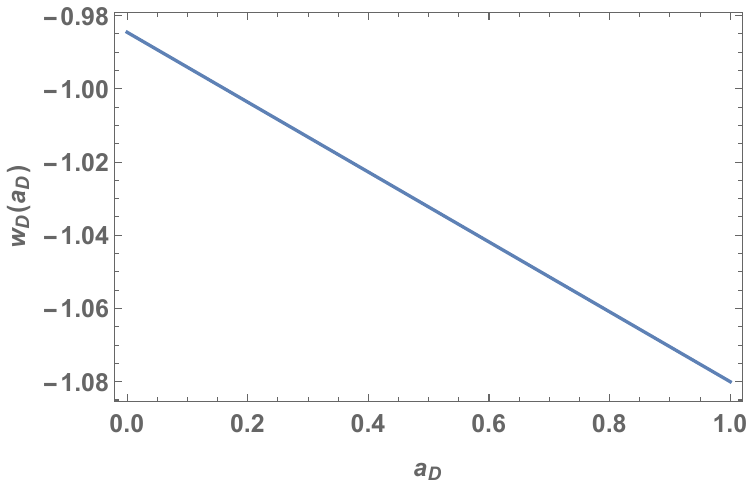}}
\end{minipage}
\hfill
\end{flushleft}

\caption{The equation of state of the effective perfect fluid on the Y-axis and volume scale factor on the X-axis. }
\label{fig:2}
\end{figure}

\begin{figure}
%\begin{tabular}{cc}
\begin{flushleft}
\begin{minipage}{0.45\linewidth}
  \centerline{\includegraphics[width=1\textwidth]{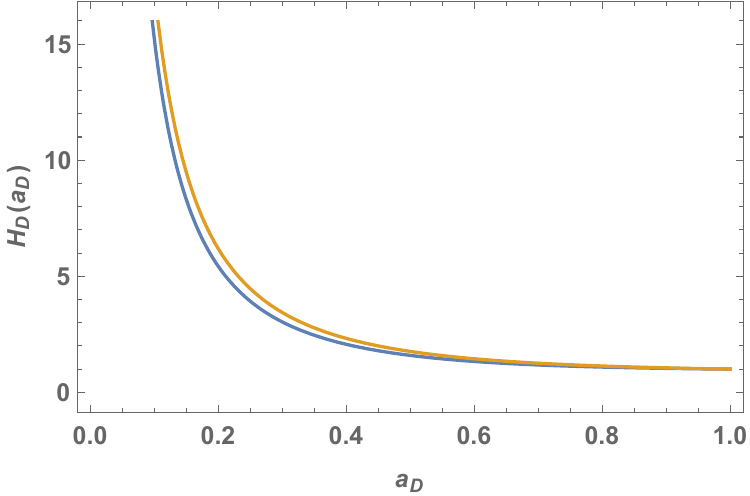}}
\end{minipage}
\hfill
\begin{minipage}{0.45\linewidth}
 \centerline{\includegraphics[width=1\textwidth]{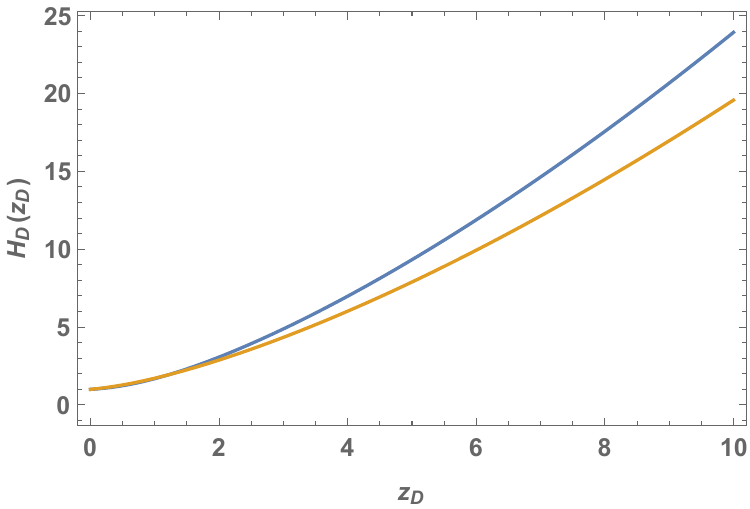}}
\end{minipage}
\end{flushleft}

\caption{Volume Hubble parameter with respect to volume scale factor on the left picture and with respect to effective redshift on the right picture. The blue line corresponding to our model and the orange line corresponding to the standard cosmology model.}
\label{fig:4}
\end{figure}

\begin{figure}
%\begin{tabular}{cc}
\begin{flushleft}
\begin{minipage}{0.45\linewidth}
  \centerline{\includegraphics[width=1\textwidth]{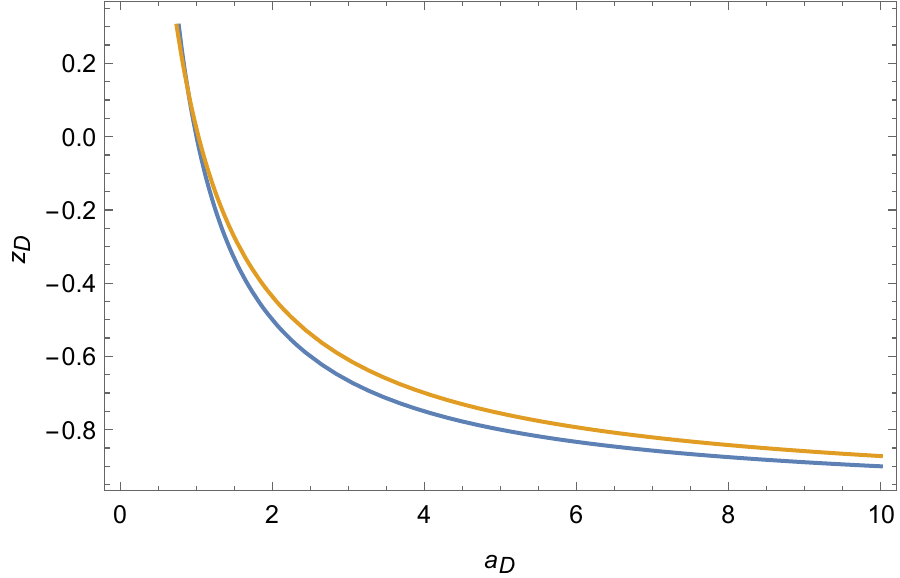}}
\end{minipage}
\hfill
\end{flushleft}

\caption{Relation between volume scale factor and effective redshift in the standard cosmology model and our model. The blue line corresponding to our model and the orange line corresponding to the standard cosmology model.}
\label{fig:5}
\end{figure}

The fitting results attained from analyzing $\chi^{2}(\Omega_{m}^{\now\CD},w_{0}^{\CD})$ by using functions Findminimum and ContourPlot in mathematica are presented in Fig.\ref{fig:1}, Table \ref{tab:2}. First of all, comparing with 0.29 \footnote{This best-fit value is obtained from a similar likelihood analysis which uses the same data sets and assumes the same prior for Hubble parameter.} in the standard cosmology model($\Lambda CDM $ model), the best-fit value of $\Omega_{m}^{\now\CD}$ in our model is smaller, this is because the template metric we use to confront the model with observations is not the same form as the FLRW metric, i.e. we assume the universe is not homogeneous-isotropic at all scale but at large-scale, at smaller scales, the universe feature strong inhomogeneities and anisotropies, and such inhomogeneities and anisotropies influence the propagation of light hence the relation between volume scale factor and effective redshift, leading to a lower values of $\Omega_{m}^{\now\CD}$ for the models to be compatible with data. Also, different from the standard cosmology model, the model we propose predicts the existence of effective dynamical dark energy, whose effective equation of state is plotted in Fig.\ref{fig:2}, which suggests that the density of the effective dynamical dark energy decreases in the distant past and becomes to increase in time when $a_{\CD}$ approach about 0.15, this increasing behavior maybe the consequence of the formation of large-scale structure. Although both the smaller $\Omega_{m}^{\now\CD}$ and late-time increasing density make the  volume Hubble parameter in the our model smaller than that in the standard cosmology model at the same volume scale factor, which is shown in the left picture of Fig.\ref{fig:4}, at the same effective redshift, However, the the situation is the opposite, as is shown in the right picture of Fig.\ref{fig:4}, this is because in these two different models, the relations between volume scale factor and effective redshift are also different, it is easy to realize that by having a look on the Fig.\ref{fig:5}.

\begin{table}
\centering
\begin{tabular}{|lcc|}
\hline
{$z$}   & $H(z)$ &  Ref.\\
\hline
$0.0708$   &  $69.0\pm19.68$         &  Zhang et al. (2014)-\cite{Zhang2014}   \\
    $0.09$       &  $69.0\pm12.0$            &  Jimenez et al. (2003)-\cite{Jimenez2003}   \\
    $0.12$       &  $68.6\pm26.2$           &  Zhang et al. (2014)-\cite{Zhang2014}  \\
    $0.17$       &  $83.0\pm8.0$             &  Simon et al. (2005)-\cite{Simon2005}     \\
    $0.179$     &  $75.0\pm4.0$           &  Moresco et al. (2012)-\cite{Moresco2012}     \\
    $0.199$     &  $75.0\pm5.0$            &  Moresco et al. (2012)-\cite{Moresco2012}     \\
    $0.20$         &  $72.9\pm29.6$         &  Zhang et al. (2014)-\cite{Zhang2014}   \\
    $0.27$       &  $77.0\pm14.0$         &    Simon et al. (2005)-\cite{Simon2005}   \\
    $0.28$       &  $88.8\pm36.6$        &  Zhang et al. (2014)-\cite{Zhang2014}   \\
    $0.352$     &  $83.0\pm14.0$          &  Moresco et al. (2012)-\cite{Moresco2012}   \\
    $0.3802$     &  $83.0\pm13.5$         &  Moresco et al. (2016)-\cite{Moresco2016}   \\
    $0.4$         &  $95\pm17.0$             &  Simon et al. (2005)-\cite{Simon2005}     \\
    $0.4004$     &  $77.0\pm10.2$          &  Moresco et al. (2016)-\cite{Moresco2016}   \\
    $0.4247$     &  $87.1\pm11.2$         &  Moresco et al. (2016)-\cite{Moresco2016}   \\
    $0.4497$     &  $92.8\pm12.9$        &  Moresco et al. (2016)-\cite{Moresco2016}   \\
    $0.4783$     &  $80.9\pm9.0$         &  Moresco et al. (2016)-\cite{Moresco2016}   \\
    $0.48$       &  $97.0\pm62.0$         &  Stern et al. (2010)-\cite{Stern2010}     \\
    $0.593$     &  $104.0\pm13.0$        &  Moresco et al. (2012)-\cite{Moresco2012}   \\
    $0.68$       &  $92.0\pm8.0$        &  Moresco et al. (2012)-\cite{Moresco2012}   \\
    $0.875$     &  $125.0\pm17.0$       &  Moresco et al. (2012)-\cite{Moresco2012}   \\
    $0.88$       &  $90.0\pm40.0$         &  Stern et al. (2010)-\cite{Stern2010}     \\
    $0.9$         &  $117.0\pm23.0$        &  Simon et al. (2005)-\cite{Simon2005}  \\
    $1.037$     &  $154.0\pm20.0$          &  Moresco et al. (2012)-\cite{Moresco2012}   \\
    $1.3$         &  $168.0\pm17.0$        &  Simon et al. (2005)-\cite{Simon2005}     \\
    $1.363$     &  $160.0\pm33.6$          &  Moresco (2015)-\cite{Moresco2015}  \\
    $1.43$       &  $177.0\pm18.0$         &  Simon et al. (2005)-\cite{Simon2005}     \\
    $1.53$       &  $140.0\pm14.0$        &  Simon et al. (2005)-\cite{Simon2005}     \\
    $1.75$       &  $202.0\pm40.0$         &  Simon et al. (2005)-\cite{Simon2005}     \\
    $1.965$     &  $186.5\pm50.4$         &   Moresco (2015)-\cite{Moresco2015}  \\
\hline
\end{tabular}
\caption{\label{tab:1} The current available OHD dataset.}
\end{table}
\begin{table}
\begin{center}
\begin{tabular}{cc|  cc }
\hline\hline
$\Omega_{m}^{\now\CD}$    &&  $0.23_{-0.04}^{+0.04}$
                     \\
$w_{0}^{\CD}$         &&  $-1.07_{-0.04}^{+0.03}$
                     \\
 \hline $\chi^{2}_{min}$  &&  $-1345.84$
                      \\
\hline\hline
\end{tabular}
\caption{\label{tab:2}Value of the $\chi^{2}_{min}$ and the fitting results of the parameters ($\Omega_{m}^{\now\CD},w_{0}^{\CD}$).}
\end{center}
\end{table}

\section{conclusion}
\label{sec:4}
In this paper, we follow the third way to handle the dark energy problem by proposing a backreaction model in which the effective dynamical dark energy's equation of state having the CPL form. By using a formula appeared in \cite{QING2013CONSTRAINTS}, we reduce a model parameter. Equipping with the template metric, we confront the model with latest type Ia supernova and Hubble parameter observations, and we find out that the small-scale inhomogeneities and anisotropies influence the evolution of the universe and the propagation of light, leading to the reduction of the value of $\Omega_{m}^{\now\CD}$, what's more, observational results show that the effective dark energy decreases in the distant past and becomes to increase in time when $a_{\CD}$ approach about 0.15. Both of these features indicate that the volume Hubble parameter in the our model smaller than that in the standard cosmology model at the same volume scale factor, although our model predicts a larger observational expanding rate.
\section*{Acknowledgments}
The paper is partially supported by the Natural Science Foundation of China.

\bibliographystyle{spphys}
\bibliography{CPL}

\end{document}